\documentclass[%
 reprint,
showpacs,preprintnumbers,
 amsmath,amssymb,
 aps,
 prb,
]{revtex4-1}

\usepackage{graphicx}
\usepackage{dcolumn}
\usepackage{bm}

\usepackage{color}

\begin{document}

\preprint{YITP}

\title{Pressure Induced Ferromagnetism in Cubic Perovskite SrFeO$_3$ and BaFeO$_3$}

\author{Zhi Li$^1$, Toshiaki Iitaka$^2$, and Takami Tohyama$^1$}
\affiliation{%
$^1$Yukawa Institute for Theoretical Physics, Kyoto University, Kyoto 606-8502, Japan\\
$^2$Computational Astrophysics Laboratory, RIKEN, 2-1 Hirosawa, Wako, Saitama 351-0198, Japan\\
}%



\date{\today}

\begin{abstract}
The spin order in cubic perovskite SrFeO$_{3}$ and BaFeO$_{3}$ under high pressure is studied by density functional theory (DFT) calculation with local spin density approximation plus Hubbard $U$ (LSDA+$U$). At ambient pressure, A-type and G-type helical spin orders are almost degenerate in BaFeO$_{3}$ whose lattice constant is 3.97~{\AA}. When the lattice constant is reduced to 3.85~{\AA} which is same as the lattice constant of SrFeO$_{3}$ at ambient pressure, G-type helical spin order becomes stable, being consistent with SrFeO$_{3}$. This is because superexchange interaction is enhanced as compared with double exchange interaction. Phase transition from helical spin state to ferromagnetic state in both SrFeO$_{3}$ and BaFeO$_{3}$ takes place if the lattice constant is further reduced to 3.70 {\AA}. This is because reduced local spin moment weakens the contribution from superexchange interaction. Our result agrees with recent experimental result of BaFeO$_{3}$ under high pressure. Additionally, our calculation predicts that half-metal BaFeO$_{3}$ at ambient pressure will become a good metal under high pressure.
\end{abstract}

\maketitle

Helical spin order in cubic perovskite $A$FeO$_3$ ($A$=Ca, Sr, Ba), where Fe$^{4+}$ is in a high spin configuration d$^{4}$, has attracted lots of research interest for their potential application in spintronic devices.~\cite{Ishiwata11,Lebon04,Adler06} All of them present helical spin order below 115~K, 134~K, and 111~K for $A$=Ca, Sr, and Ba, respectively.~\cite{Kawasaki98,Woodward00,Takeda72,Hayashi11} The Fe3$d$ electrons in these materials can be divided into two classes: conducting and localized electrons. Three localized electrons occupy t$_{2g}$ orbitals and one electron occupies double degenerated e$_{g}$ orbitals. The interaction between conducting electron and localized electron is described by Hund coupling. In addition, charge-transfer energy $\Delta$ defined as the energy cost to move an electron from oxygen 2$p$ orbital to Fe3$d$ orbitals shows a negative value,~\cite{Maekawa04,Bocuet92, AEB92} implying that metallic conduction mainly occurs on oxygen band.  The helical spin order can be understood from the competition of double exchange (DE) and superexchange (SE) interactions, the former and the latter of which favors ferromagnetism (FM) and anti-ferromagnetism (AFM), respectively.~\cite{Mostovoy05, ZL12}

It is well-known that electronic structure is tunable under high pressure. Phase transition from helical spin state to FM state in SrFeO$_3$ under the pressure of 7~GPa has been reported.~\cite{Kawakami05}  Very recently, the evolution of spin order in BaFeO$_3$ under pressure has been studied.~\cite{Kawakami12} At ambient pressure, BaFeO$_3$ shows helical spin order, which changes to FM under very weak external magnetic field, $\sim$0.3~T.~\cite{Hayashi11} The helical spin order is stabilized with increasing pressure, but FM finally becomes stable under pressure above 30~GPa. There is no structural phase transition, since the cubic symmetry of BaFeO$_3$ preserves up to 50~GPa. The electrical resistance decreases under high pressure.

Hydrostatic pressure $P$ reduces the lattice constant $a$ in $A$FeO$_3$. The compression of $a$ leads to the increase of the hopping integral $pd\sigma$ representing the hybridization between O2$p$ and Fe3$d$ orbitals. Since the DE and SE energies are roughly proportional to $pd\sigma$ (ref.~11) and $(pd\sigma)^4$, respectively, increasing $pd\sigma$ gives the enhancement of SE as compared with DE. This explains the stabilization of helical spin order under pressure in BaFeO$_3$, in which the G- and A-type helical spin orders are almost degenerate at ambient pressure.\cite{ZL12} However, this cannot explain the stabilization of FM under further pressure. Therefore, it is significant to study the electronic structure of SrFeO$_3$ and BaFeO$_3$ under high pressure, which is helpful to understand the FM phase transition and the behavior of electrical resistance.

In this paper, we perform density functional theory (DFT) calculations with local spin density approximation plus Hubbard $U$ (LSDA+$U$) for both SrFeO$_{3}$ and BaFeO$_{3}$. At ambient pressure, the lattice constant of SrFeO$_{3}$ and BaFeO$_{3}$ is 3.85 {\AA} and 3.97 {\AA}, respectively, and both materials show helical spin order. We find that BaFeO$_{3}$ represents stable G-type helical spin order that is the same order as SrFeO$_3$, if the lattice constant is reduced to $a=3.85$~{\AA}. We also find that FM is stable in both SrFeO$_{3}$ and BaFeO$_{3}$ if $a$  is further reduced to 3.70~{\AA}. Our result for BaFeO$_{3}$ agrees with recent experimental results under high pressure.~\cite{Kawakami12} According to the DE model including oxygen orbitals and SE,~\cite{Mostovoy05} the increase of $pd\sigma$ stabilizes FM when SE is unchanged. The SE energy is given by the product of neighboring spin moment with coupling parameter $J_\mathrm{SE}$. With decreasing $a$, $J_\mathrm{SE}$ increases according to a relation $J_\mathrm{SE}\propto (pd\sigma)^4$, while the product of neighboring spin moment decreases because of the suppression of spin moment on Fe site. As a result, the SE energy may become less sensitive to $a$ under high pressure. This view can explain the stabilization of FM obtained by our first-principles calculations.

Non-collinear spin polarized calculations are performed by VASP \cite{VASP} within LSDA+$U$~\cite{Liechtenstein} in a primitive cell with 10$\times$10$\times$10 k-point grid. For the non-collinear spin order calculation, the wave function is in the form of spinor and a generalized Bloch boundary condition is adopted.~\cite{Sandratskii}.  The projector augmented wave pseudopotentials with Ceperly-Alder exchange-correlation with 500~eV energy cutoff is used in our calculation. The energy resolution is set to be 0.01~meV per unit cell. For band dispersions and density of states (DOS), 40$\times$40$\times$40 k-point grid is adopted for their self-consistent calculations.
\begin{figure}[t]
\includegraphics[width=7cm]{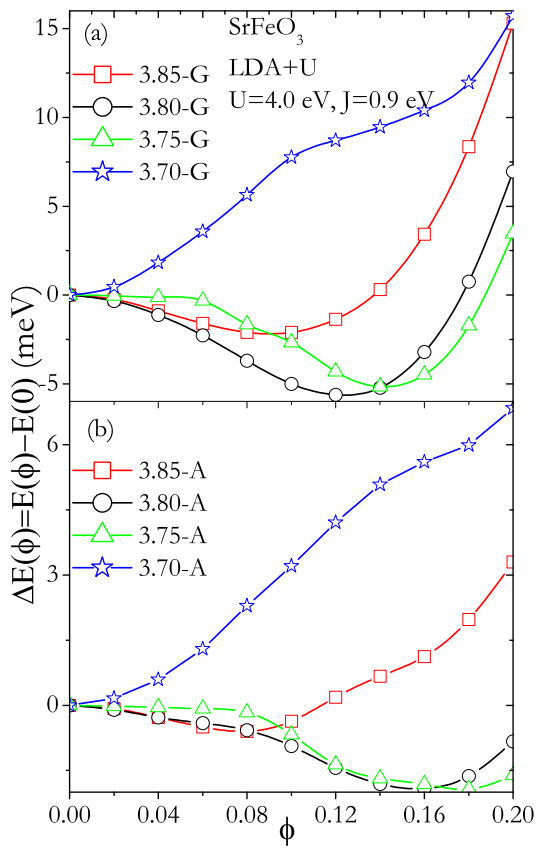}
\caption{\label{fig1} (Color online) The $\phi$ dependence of the total energy difference per unit cell, $\Delta E(\phi)\equiv E(\phi)-E(\phi=0)$, obtained by LSDA+$U$ for various lattice constant $a$ shown as numbers in SrFeO$_3$. (a) G-type helical spin order; (b) A-type helical spin order. At ambient pressure, $a$=3.85~\AA. For the definition of $\phi$, see text. }
\end{figure}

\begin{figure}[t]
\includegraphics[width=7cm]{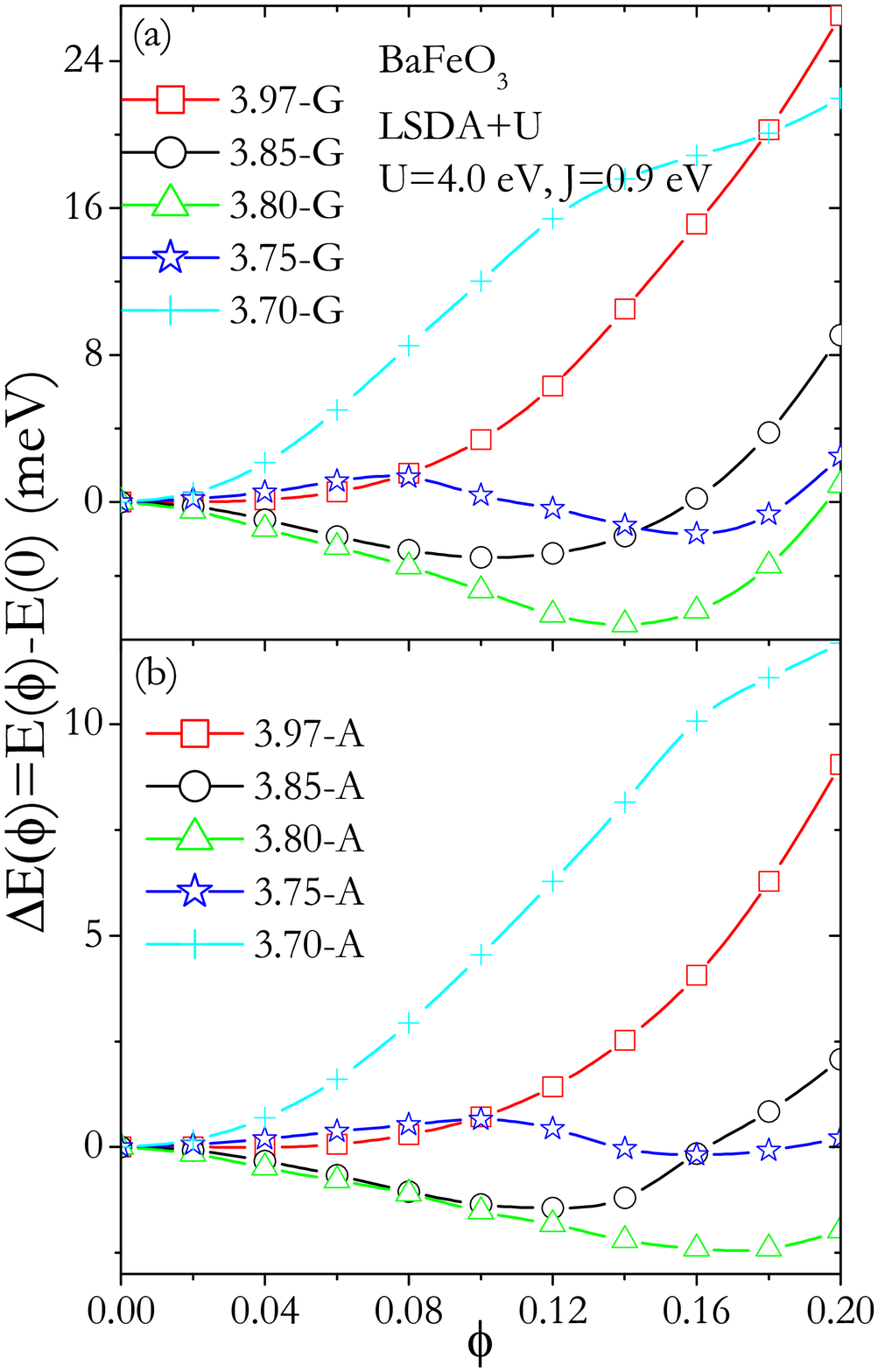}
\caption{\label{fig2} (Color online) The $\phi$ dependence of the total energy difference per unit cell, $\Delta E(\phi)\equiv E(\phi)-E(\phi=0)$, obtained by LSDA+$U$ for various lattice constant $a$ shown as numbers in BaFeO$_3$. (a) G-type helical spin order; (b) A-type helical spin order. At ambient pressure, $a$=3.97~\AA. For the definition of $\phi$, see text. }
\end{figure}

\begin{figure}[t]
\includegraphics[width=7cm]{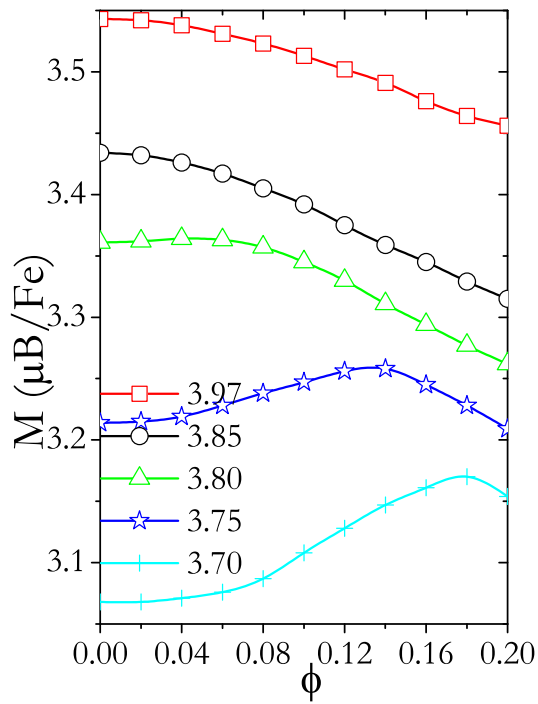}
\caption{\label{fig3} (Color online) The spin moment $M$ in BaFeO$_3$ with G-type helical spin order obtained by LSDA+$U$ as a function of $\phi$ for several values of $a$.}
\end{figure}

 It is necessary to add on-site Coulomb interaction $U$ and exchange interaction $J$ to LSDA in order to obtain a helical spin order.~\cite{ZL12}  In the present study, we take a set of parameters, $U=4.0$~eV and $J=0.9$~eV, and for simplicity, we ignore the $a$ dependence of $U$ and $J$. These parameters are slightly larger than those used in a previous paper ($U=3.0$~eV and $J=0.6$~eV).~\cite{ZL12} However, similar to the previous study, this set also gives a propagating vector of helical spin order that is consistent with the observed one in SrFeO$_{3}$ at ambient pressure. Furthermore, for BaFeO$_{3}$ a calculated lattice constant where the FM transition occurs agrees well with an observed one as will be shown below. We emphasize that the physics discussed in the present paper is independent of the choice of the parameters.

The propagating vector of A-type and G-type helical spin orders is defined as $\vec{q}=\frac{2\pi}{a}(\phi,0,0)$ and $\vec{q}=\frac{2\pi}{a}(\phi,\phi,\phi)$, respectively. To find out the optimal value of $\phi$, we define the $\phi$-dependent total energy measured with respect to the energy of FM state, $\Delta E(\phi)\equiv E(\phi)-E(\phi=0)$. The energy in SrFeO$_{3}$ with different lattice constant is shown in Fig.~\ref{fig1}. At ambient pressure ($a$=3.85~\AA), the G-type order with $\phi$=0.09 shows an energy minimum. This $\phi$ is close to an observed one ($\phi$=0.112).~\cite{Takeda72} With decreasing $a$, the G-type order becomes more stable, but with further decreasing $a$ a transition from helical spin state to FM state occures irrespective of the type of order, i.e., the $\phi$ at minimum energy becomes zero when $a$ changes from 3.75~{\AA} to 3.70~{\AA} in both the A- and G-type orders. This result is qualitatively consistent with the experimental results of a transition from AFM to FM induced by pressure.~\cite{Kawakami05}. An observed $a$ at the transition pressure (7~GPa) is about $a$=3.78~\AA~\cite{Kawakami05,Kawakami03}, which is slightly larger than the calculated one. The difference may partly come from the choice of $U$ and $J$. We note that smaller $U$ and $J$ gives a larger $a$ at the transition.

The $\Delta E(\phi)$ in BaFeO$_{3}$ with A-type and G-type helical spin orders is shown in Fig.~\ref{fig2}(a) and Fig.~\ref{fig2}(b), respectively. At ambient pressure ($a$=3.97~{\AA}), the A-type and G-type orders are almost degenerate with small propagating vector.~\cite{ZL12} With increasing pressure, i.e., reducing $a$ from 3.97~{\AA} at ambient pressure to $a$=3.85~{\AA}, both the A-type and G-type orders become more stable with larger $\phi$. We note that at $a$=3.85~{\AA} the G-type order is more stable than the A-type order. The value of $\phi$ at minimal $\Delta E(\phi)$ with $a$=3.85~{\AA} is almost the same as that in SrFeO$_{3}$ with the same $a$. As the case of SrFeO$_3$, BaFeO$_{3}$ shows FM if $a$ is further reduced (see Figs.~\ref{fig2}(a) and (b)). The pressure induced FM is thus common among $A$FeO$_3$.

\begin{figure}[t]
\includegraphics[width=6cm]{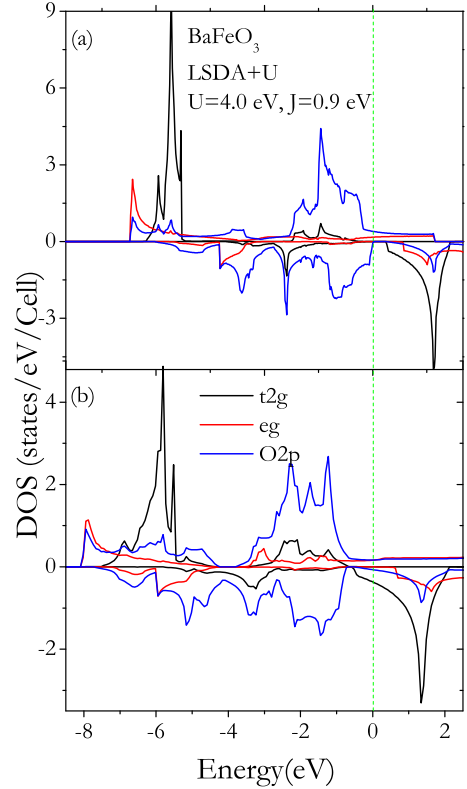}
\caption{\label{fig4} (Color online) Density of states (DOS) of BaFeO$_3$ in the FM state calculated by LSDA+$U$ with $U=4.0$~eV and $J=0.9$~eV. The positive side of the DOS denotes the up-spin DOS, while the negative side denotes the down-spin DOS. Dotted vertical line at zero energy represents the Fermi level. Black solid line and red solid lines represent the DOS of Fe$t_{2g}$ and Fe$e_{g}$ orbitals, respectively. The partial DOS of O2$p$ orbitals is shown by solid blue line. (a) $a$=3.97~\AA (ambient pressure). (b) $a$= 3.70~\AA (under pressure).}
\end{figure}

\begin{figure}[t]
\includegraphics[width=7cm]{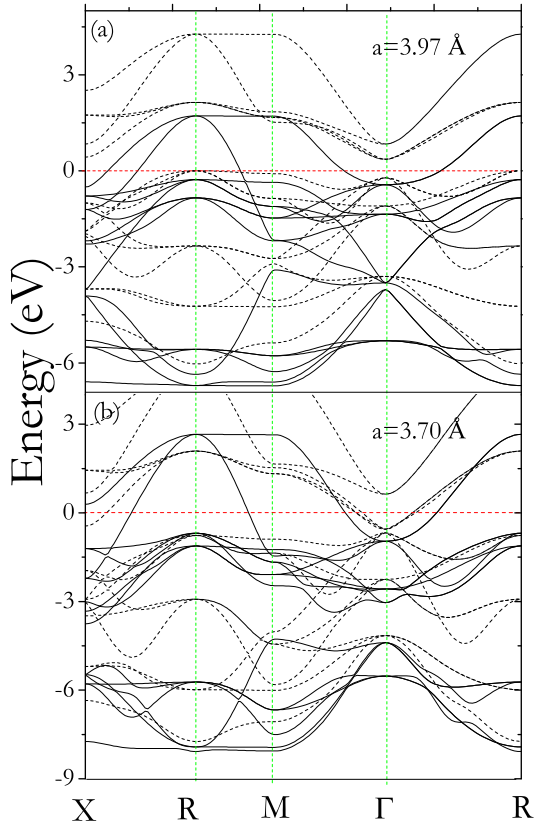}
\caption{\label{fig5} (Color online) Band structure of ferromagnetic BaFeO$_3$ calculated by LSDA+$U$. Solid and dotted lines represent spin-up and spin-down bands, respectively. (a) $a$=3.97~{\AA} and (b) $a$=3.70~{\AA}.}
\end{figure}

The energy minimum of G-type helical spin order in BaFeO$_{3}$ (see Fig.~\ref{fig2}(b)) shifts to the right-hand side when we reduce $a$ from 3.97~{\AA} to 3.75~{\AA}. The shift of the energy minimum can be explained by the competition between DE and SE interactions. The reduction of $a$ leads to the increase of $pd\sigma$. Since the DE and SE energies are roughly proportional to $pd\sigma$ (ref.~11) and $(pd\sigma)^4$, respectively, the increase of $pd\sigma$ gives the enhancement of SE interaction, comparing with DE. The enhanced SE stabilizes helical spin order at large $\phi$.

Figure~\ref{fig3} shows the calculated spin moment $M$ as a function of $\phi$ for several values of $a$ in BaFeO$_{3}$ with G-type helical spin order. The magnitude of $M$ decreases with decreasing $a$, and $\phi_\mathrm{m}$, defined as the value of $\phi$ where $M$ shows maximum, shifts toward larger $\phi$. We find that the shift of $\phi_\mathrm{m}$ is consistent with the shift of the minimal $\Delta E(\phi)$ in Fig.~\ref{fig2}(b) when 3.97~{\AA} $\leq a \leq$3.75~{\AA}. This is a reasonable behavior, since it is expected that the energy is gained if the magnetization is large.

With further reducing $a$ to 3.70~{\AA} in Fig.~\ref{fig2}(a), the global energy minimum shifts to a FM state. This means that DE finally comes over SE. We still can see the trace of a local minimum around $\phi$=0.18. Correspondingly, the maximum of $M$ is located around $\phi$=0.18. It is interesting to notice that the transition to the FM state seems to be discontinuous, i.e., a first-order transition. This is clear from the $E(\phi)$ curve for $a$=3.75~\AA, where there appear two minima at $\phi$=0 and 0.16. If the DE interaction is only between nearest-neighboring sites, $E(\phi)$ has only a single minimum. Therefore, the discontinuous transition indicates the presence of longer-range DE interactions. The change from helical to FM state between $a$=3.75~{\AA} and $a$=3.70~{\AA} obtained by our calculation agrees with the emergence of FM state in BaFeO$_{3}$ at $a$=3.73~{\AA} observed by experiment~\cite{Kawakami12}.

Now, a question is why the SE will fail to stabilize helical spin order if the lattice constant reduced to $a$=3.70~{\AA}. The SE energy between nearest-neighbor localized spins is given by $E_\mathrm{SE}=J_\mathrm{SE}\sum_{<i,j>}\left< \mathbf{S}_i\cdot\mathbf{S}_j \right>$, where $J_\mathrm{SE}$ is the SE coupling parameter, $\mathbf{S}_i$ is the spin operator at site $i$, $<i,j>$ runs over the nearest-neighbor pairs, and $\left< \cdots \right>$ represents an average of an static quantity. The SE energy is, thus, proportional to the product of neighboring spin moments. Since $M$ decreases with decreasing $a$ as shown in Fig.~\ref{fig3}, $\left< \mathbf{S}_i\cdot\mathbf{S}_j \right>$ is also expected to decrease. This will compensate the effect of the increase of $J_\mathrm{SE}\propto (pd\sigma)^4$, and thus the SE energy will become less $a$-dependent. As a consequence, the effect of DE overcomes the effect of SE, leading to the first-order transition to FM around $a$=3.70~{\AA}.

To understand the mechanism of the reduction of $M$ under high pressure, it is convenient to study DOS in BaFeO$_{3}$. Here, we focus on the cases with $a$=3.97~{\AA} and 3.70~{\AA}, both of which show the energy minimum near or at $\phi$=0. Comparing DOS at $a$=3.97~{\AA} with that at $a$= 3.70~{\AA} shown in Fig.~\ref{fig4}(a) and Fig.~\ref{fig4}(b), we clearly see the increase of the energy width of DOS under pressure.  We note that the increase of band width usually leads to smaller $M$ through the suppression of DOS in the paramagnetic phase.

The increase of band width also induces the change of DOS at the Fermi level. At $a$=3.97~{\AA}, only electrons with up spin contribute to the conductivity, and the DOS of down spin is almost zero at the Fermi level as shown in Fig.~\ref{fig4}(a). This means that ferromagnetic BaFeO$_3$ is half-metallic at ambient pressure.~\cite{ZL12}  At $a$=3.70~{\AA}, DOS at the Fermi level becomes nonzero and small part of t$_{2g}$ DOS is occupied by down-spin electrons. This is consistent with the decrease of $M$. The band dispersions of BaFeO$_{3}$ in the FM state are shown in Fig.~\ref{fig5}. Comparing the dispersions for $a$=3.97~{\AA} and $a$=3.70~{\AA}, we find that main changes under high pressure take place at the $\Gamma$ and X points. At the $\Gamma$ point, several spin-down bands descend and cross the Fermi level.  At X point, one spin-up band lifts and one spin-down band descends. The downward shift of the spin-down bands is consistent with the emergence of extra $t_{2g}$ electrons near the Fermi level in the down-spin DOS shown in Fig.~\ref{fig4}(b). In other words, a half metallic nature in BaFeO$_{3}$ at ambient pressure is lost under high pressure and becomes a good metal. This explains the observed suppression of electric resistance under high pressure.~\cite{Kawakami12}

In summary, the spin order of SrFeO$_{3}$ and BaFeO$_{3}$ under high pressure is studied by DFT calculation with LSDA+$U$. A transition from helical spin state to FM state in SrFeO$_{3}$ and BaFeO$_{3}$ under high pressure is reproduced. The transition is of the first-order type. The mechanism of the transition is associated with the evolution of DE energy and SE energy. The DE energy increases with decreasing $a$ because of the enhancement of hopping integral $pd\sigma$. This favors FM. However, the gain of the SE energy competes with the DE energy, and as a result, the helical spin order is stabilized in the range of $a$=3.97$\sim$3.75~{\AA}. With further reducing $a$, the effect of SE fades out because of reduced local spin moment on Fe. Our calculated results agree with the recent experimental result on BaFeO$_{3}$. From the $a$ dependence of DOS in the FM state, we predict that a half metallic character of ferromagnetic BaFeO$_{3}$ at ambient pressure is lost under high pressure.

We would like to thank T. Kawakami for providing us unpublished data. Z.L. is grateful to the global COE program of "Next Generation Physics, Spun from Universality and Emergence". This work was also supported by the Strategic Programs for Innovative Research (SPIRE), the Computational Materials Science Initiative (CMSI), a Grant-in-Aid for Scientific Research (Grant No. 22340097) from MEXT, and the Yukawa International Program for Quark-Hadron Sciences at YITP, Kyoto University. Part of the numerical calculations was performed in the supercomputing facilities in YITP, Kyoto University, and RICC in RIKEN.



\begin{thebibliography}{99}
\bibitem{Ishiwata11} S. Ishiwata, M. Tokunaga, Y. Kaneko, D. Okuyama, Y. Tokunaga, S. Wakimoto, K. Kakurai, T. Arima, Y. Taguchi, and Y. Tokura, Phys. Rev. B {\bf 84}, 054427 (2011).

\bibitem{Lebon04} A. Lebon, P. Adler, C. Bernhard, A.V. Boris, A.V. Pimenov, A. Maljuk, C.T. Lin, C. Ulrich, and B. Keimer, Phys. Rev. Lett. {\bf 92}, 037202 (2004).

\bibitem{Adler06} P. Adler, A. Lebon, V. Damljanovic, C. Ulrich, C. Bernhard, A. V. Boris, A. Maljuk, C. T. Lin, and B. Keimer, Phys. Rev. B {\bf 73}, 094451 (2006).

\bibitem{Kawasaki98} S. Kawasaki, M. Takano, R. Kanno, T. Takeda, and A. Fujimori, J. Phys. Soc. Jpn. {\bf 67}, 1529 (1998).

\bibitem{Woodward00} P. M. Woodward, D. E. Cox, E. Moshopoulou, A. W. Sleight, and S. Morimoto, Phys. Rev. B {\bf 62}, 844 (2000).

\bibitem{Takeda72} T. Takeda, Y. Yamaguchi, and H. Watanabe, J. Phys. Soc. Jpn. {\bf 33}, 967 (1972).

\bibitem{Hayashi11} N. Hayashi, T. Yamamoto, H. Kageyama, M. Nishi, Y. Watanabe, T. Kawakami, Y. Matsushita, A. Fujimori, and M. Takano, Angew. Chem. Int. Ed. {\bf 50}, 12547 (2011).

\bibitem{Maekawa04} S. Maekawa, T. Tohyama, S. E. Barnes, S. Ishihara, W. Koshibae, and G. Khaliullin, Physics of Transition Metal Oxides, Springer Series in Solid State Sciences Vol. 144 (Springer-Verlag, Heidelberg, 2004).

\bibitem{Bocuet92} A. E. Bocquet, A. Fujimori, T. Mizokawa, T. Saitoh, H. Namatame, S. Suga, N. Kimizuka, Y. Takeda, and M. Takano, Phys. Rev. B {\bf 45}, 1561 (1992).

\bibitem{AEB92} A. E. Bocquet, T. Mizokawa, T. Saitoh, H. Namatame, and A. Fujimori, Phys. Rev. B {\bf 46}, 3771(1992).

\bibitem{Mostovoy05} M. Mostovoy, Phys. Rev. Lett. {\bf 94}, 137205 (2005).

\bibitem{ZL12} Z. Li, R. Laskowski, T. Iitaka, and T. Tohyama, Phys. Rev. B {\bf 85}, 134419 (2012).

\bibitem{Kawakami05} T. Kawakami and S. Nasu, J. Phys.: Condens. Matter {\bf 17},  S789 (2005).

\bibitem{Kawakami12}T. Kawakami, H. Tanaka, Y. Watanabe, A. Kawasaki, Y. Nakakura, T. Kamatani, N. Idegomori, N. Hayashi, S. Nasu, and M. Takano, private communications.

\bibitem{Kawakami03}T. Kawakami, S. Nasu, K. Kuzushita, T. Sakai, S. Morimoto, T. Yamada, S. Endo, S. Kawasaki, and M. Takano, J. Phys. Soc. Jpn. {\bf 72}, 33 (2003).

\bibitem{VASP} D. Hobbs, G. Kresse, and J. Hafner, Phys. Rev. B 62, 11556 (2000).

\bibitem{Liechtenstein} A. I. Liechtenstein, V. I. Anisimov, and J. Zaanen, Phys. Rev. B {\bf 52}, R5467 (1995)

\bibitem{Sandratskii} L. M. Sandratskii, Phys. Status Solidi B 136, 167 (1986).


\end{thebibliography}

\end{document}